\documentclass[conference]{IEEEtran}

\usepackage{amsfonts, amsmath, amssymb, epsfig, graphicx}
\usepackage{epstopdf}
\usepackage{mathtools}
\usepackage{epsfig}
\usepackage{times}
\usepackage{float}
\usepackage{afterpage}
\usepackage{amsmath}
\usepackage{amstext}
\usepackage{amssymb,bm}
\usepackage{latexsym}
\usepackage{color}
\usepackage{graphicx}
\usepackage{amsmath}
\usepackage{amsthm}
\usepackage{graphicx}
\usepackage{pstricks}
\usepackage{subfigure}
\usepackage{booktabs}
\usepackage{enumerate}
\usepackage[top=1in, bottom=0.75in, left=0.75in, right=0.75in]{geometry}

\newtheorem{theorem}{Theorem}

\ifCLASSINFOpdf
\else
\fi
\begin{document}

\title{The DoF of the Asymmetric MIMO Interference Channel with Square Direct Link Channel Matrices}

\author{
\IEEEauthorblockN{Tang Liu$^{\dagger}$, Daniela Tuninetti$^{\dagger}$, Syed A. Jafar$^*$}
$^{\dagger}$University of Illinois at Chicago,
Chicago, IL 60607, USA, 
Email: {tliu44, danielat}@uic.edu\\
$^*$University of California at Irvine,
Irvine, CA 92697, USA,   
Email: syed@uci.edu}
\maketitle
\begin{abstract}
This paper studies the sum Degrees of Freedom (DoF) of $K$-user {\em asymmetric} MIMO Interference Channel (IC) with square direct link channel matrices, that is, the $u$-th transmitter and its intended receiver have $M_u\in\mathbb{N}$ antennas each, where $M_u$ need not be the same for all $u\in[1:K]$. 

Starting from a $3$-user example, it is shown that existing cooperation-based outer bounds are insufficient to characterize the DoF. Moreover, it is shown that two distinct operating regimes exist. With a {\it dominant} user, i.e., a user that has more antennas than the other two users combined, 
it is DoF optimal to let that user transmit alone on the IC.
Otherwise, it is DoF optimal to {\em decompose} and operate the 3-user MIMO IC as an $(M_1+ M_2+M_3)$-user SISO IC. This indicates that MIMO operations are useless from a DoF perspective in systems without a dominant user.

The main contribution of the paper is the derivation of a novel outer bound for the general $K$-user case that is tight in the regime where a dominant user is not present; this is done by generalizing the insights from the 3-user example to an arbitrary number of users. 
\end{abstract}


\section{Introduction}\label{sec:intro}
{
Interference channels (IC) have been extensively studied in the past years due to their practical relevance.
The capacity of even the simple two-user case is still open in general.
For the Gaussian noise IC progress has been made by focusing on the degrees-of-freedom (DoF), or scaling of the sum-capacity with signal-noise-ratio (SNR) as SNR grows to infinity. A signaling scheme, known as interference alignment \cite{IA-K-user}, has been shown to achieve 1/2 the interference-free capacity for each user for almost all channel realizations, regardless of the number of users, in single antenna systems. This showed the surprising result that ICs are not intrinsically interference limited.

Multiple-input-multiple-output (MIMO) techniques are widely used in practical wireless communication systems as a means to increase the spectral efficiency. The complete characterization of the DoF of a general multiuser MIMO IC has been elusive so far.
The case where every node has the same number of antennas was solved in \cite{IA-K-user}, where it was shown that MIMO operations are not needed to achieve the optimal DoF. 
The question whether the same remains true in {\it asymmetric} MIMO IC has been answered in some special cases only. 

In \cite{DoF-MIMO} Jafar and Fakhereddin fully characterized the DoF of the 2-user MIMO IC with arbitrary number of antennas at each node. Their result has served as a fundamental outer bound for the $K$-user MIMO IC where each transmitter has $M$ antennas, each receiver has $N$ antennas, and $M\not=N$, indicated as the $(M\times N)^K$ IC \cite{MNKuser,IA-k-user-MIMO,alignment-chain,3usergeo,squaregrass,kuser-grass}. The idea is to partition both the set of transmitters and the set of receivers into two groups, let the users in each group perfectly cooperate and thus outer bound the performance of the original IC by that of the so obtained 2-(super)user IC.
For the $(M\times N)^K$ IC, MIMO operations are needed in order to attain the optimal DoF; however it was observed that, except for some values of $M/N$, either $M$ or $N$ can be reduced without affecting the DoF \cite{MNKuser}, \cite{alignment-chain}.
For this $(M\times N)^K$ model, both the achievability and converse proofs relied on the  the symmetry of antennas across users and it is not a priori clear how to generalize them to settings that lack this symmetry.

The case where $K$ MIMO users share the same channel and each node can have different number of antennas has not received so much attention as of yet, to the best of our knowledge. The reason may lie in the fact that known bounds for ``almost symmetric'' ICs do not seem to be tight in the general case. 
In this work we study the class of general asymmetric MIMO ICs with {\it square direct link channel matrices}, that is, each transmitter and its corresponding receiver have the same number of antennas, but different transmitter-receiver pairs can have different number of antennas. Although this setting is not fully general yet, it is a first step towards understanding the impact of heterogeneous devices in ad-hoc networks.

The main contribution of the paper is a full DoF characterization for the proposed setting.
First we show that existing cooperation-based outer bounds are insufficient to characterize the DoF and derive a novel DoF outer bound. The novel bound reveals that two distinct operating regimes exist. With a {\it dominant} user, i.e., a user that has more antennas than all the other users combined, it is optimal to let that user transmit alone on the IC.
Otherwise, it is optimal to {\em decompose} and operate the MIMO IC as a multiuser single-input-single-output (SISO) IC where the number of users is given by the total number of transmit antennas. This rather surprising result indicates that MIMO operations are useless from a DoF perspective in systems without a dominant user if the direct link channel matrices are square.
}

The paper is organized as follows. 
Section \ref{sec:model} presents the channel model and summarizes known bounds.
Section \ref{sec:example} highlights the main ingredients in the converse proof by means of a simple 3-user example.
The rigorous proof for the general $K$-user case is provided in Section \ref{sec:Kuser}. 
Section \ref{sec:conclusion} concludes the paper.

\section{Channel Model and Known Bounds}
\label{sec:model}

\subsection{Channel model}
\label{sec:model:ch}
We consider a specific multiuser {\em asymmetric} MIMO IC that consists of $K$ transmitter-receiver pairs sharing the same wireless channel and thus interfering with one another.  We let $M_u$ be the number of antennas at ${\rm Tx}_u$ and at ${\rm Rx}_u$, $u\in[1:K]$, where without loss of generality  $M_1 \geq M_2 \geq \ldots \geq M_K \geq 1$.
The channel input-output relationship is 
\begin{subequations} 
\begin{align}
\bar{Y}_i & = \sum_{k\in [1:K]} \bar{H}_{ik} \bar{X}_k +\bar{Z}_i \in\mathbb{C}^{M_i \times 1}, \ i\in [1:K],
\\
& \bar{X}_i\in\mathbb{C}^{M_i \times 1} : \mathbb{E}[\lVert\bar{X}_i\rVert^2] \leq \mathsf{P},
\\
& \bar{Z}_i\in\mathbb{C}^{M_i \times 1} : \bar{Z}_i \sim \mathcal{N}(\mathbf{0},\mathbf{I}),
\end{align} \label{eq:ch model general}%
\end{subequations} 
where $\bar{H}_{ij}\in\mathbb{C}^{M_i \times M_j}$ is the channel matrix from ${\rm Tx}_j$ to ${\rm Rx}_i$, $(i,j)\in [1:K]^2$. 
${\rm Tx}_i$ has a message $W_i$, of rate $R_i(\mathsf{P})$, where $\mathsf{P}$ is the transmit power, for ${\rm Rx}_i, \ i\in [1:K]$.
Achievable rates and capacity region are defined in the usual way \cite{NIT}.

In this work we are interested in the high-SNR regime, i.e., $\mathsf{P}\gg1$, and will use the DoF as performance metric.
The (sum) DoF $d_\Sigma$ is defined as 
\begin{align}
d_\Sigma :=  \sup \sum_{i\in [1:K]}d_{i}, \
\end{align}
where the supremum is over all achievable rate vectors $(R_1(\mathsf{P}),\ldots,R_K(\mathsf{P}))$ and where $d_i$ is the DoF of $i$-th user defined as
$d_i := \lim_{\mathsf{P}\to+\infty}\frac{R_i(\mathsf{P})}{\log(\mathsf{P})}$ for $i\in [1:K].$

\subsection{Inner bound}
\label{sec:model:inner}
An achievable scheme is as follows.
By `disabling' MIMO operations, i.e., treating each pair of antennas as a separate user, we can transform the MIMO IC into a SISO IC with $\sum_{i\in [1:K]}M_i$ users; by interference alignment we can achieve $1/2$ DoF per user \cite{IA-K-user,real-alignment-single}. We shall refer to this simple achievable scheme as the {\em decomposition} inner bound \cite{decomposition}.

Another simple achievable scheme is to let only ${\rm Tx}_1$  (the user with the largest number of antennas) transmit, and achieve 
$d_1 = M_1,  \  d_i=0,\ i \in[2:K]$.

By combining these two schemes, the DoF of our asymmetric MIMO IC satisfies
\begin{align}
\max \left(M_1, \frac{\sum_{i\in [1:K]}M_i}{2} \right) \leq d_\Sigma.
\label{eq:sumDoF in}
\end{align}

\subsection{Outer bound}
\label{sec:model:upper}
The DoF of 2-user MIMO IC with arbitrary number of antennas at each node was derived in \cite{DoF-MIMO}. This result is widely used in DoF converse proofs (see for example \cite{MNKuser,IA-k-user-MIMO}) where the main idea is to reduce a $K$-user MIMO IC to a 2-user one by either `silencing' all but two users, or by using cooperation to obtain a 2-user MIMO IC.
Therefore, by partitioning the $K$ users into two groups so as to form two `super users' and by applying the result of \cite{DoF-MIMO}, we immediately obtain that the DoF of our asymmetric MIMO IC satisfies
\begin{align}
  d_\Sigma \leq \min_{\mathcal{S}\subseteq[1:K]} \max\left(\sum_{i\in \mathcal{S}} M_i, \sum_{i\in \mathcal{S}^c} M_i \right),
\label{eq:sumDoF out}
\end{align}
where $\mathcal{S}^c$ is the complement of $\mathcal{S}$ in $[1:K]$.
We shall refer to this bound as the {\em cooperation} outer bound.

\subsection{Systems with a dominant user} 
\label{sec:model:dom}
When one user has more antennas than all the other users combined, i.e., 
\begin{align}
M_1 \geq \sum_{i\in [2:K]}M_i,
\label{eq:sumDoF dominant-user case}
\end{align}
we say that the IC has a {\it dominant} user (user~1 in our channel setting). In this case the left hand side of~\eqref{eq:sumDoF in} and the right hand side of~\eqref{eq:sumDoF out} coincide, and thus the DoF is completely characterized.
Therefore, for systems with a dominant user, the cooperation outer bound is tight and is achieved by letting only the dominant user transmit.

\subsection{Systems without a dominant user}
\label{sec:model:NOTdom}
When there is no dominant user, the inner bound in~\eqref{eq:sumDoF in} and the outer bound in~\eqref{eq:sumDoF out} do not coincide in general unless there exists a set $\mathcal{S}\subseteq[1:K]$ such that
\begin{align*}
  \sum_{i\in \mathcal{S}} M_i = \sum_{i\in \mathcal{S}^c}M_i,
\end{align*}
in which case the decomposition inner bound matches the cooperative outer bound. 
So in general either the cooperative outer bound or the decomposition inner bound is not tight.

In order to understand which bound might be loose, 
we next consider a specific 3-user IC example. Through this example we will show that the decomposition inner bound in~\eqref{eq:sumDoF in} is tight. This will provide the necessary intuition for the extension of the proof to the general $K$-user case in Section~\ref{sec:Kuser}.

\section{Example: the $(M_1,M_2,M_3)=(2,2,1)$ case}
\label{sec:example}

For the case $(M_1,M_2,M_3)=(2,2,1)$, the outer bound in~\eqref{eq:sumDoF out} gives $d_\Sigma \leq 3$ while the inner bound in~\eqref{eq:sumDoF in} gives $d_\Sigma \geq 5/2$. In this section we aim to demonstrate that the outer bound is loose. Intuitively, 3~DoF appears to be too optimistic since it is well known that the 3-user MIMO IC with $(M_1,M_2,M_3)=(2,2,2)$ has 3~DoF \cite{IA-k-user-MIMO}. Therefore, if the outer bound were tight, it would indicate that removing one antenna at each terminal of the the third transmitter-receiver pair does not impact the DoF. Cases of `antenna redundancy' are known in \cite{MNKuser}, \cite{alignment-chain}, but we shall show that this is not the case for our asymmetric MIMO IC when no dominant user exists.

In Section~\ref{sec:example:trans} we start by transforming the IC in~\eqref{eq:ch model general} into an equivalent IC in which the channel matrices contain zeros in carefully chosen positions.
In Section~\ref{sec:example:outLin} we give a `dimension counting argument' to show that no more than $5/2$~DoFs are achievable in the equivalent IC.
Finally in Section~\ref{sec:example:outIT}, we give an information theoretic proof of this intuitive argument and show the outer bound $d_\Sigma \leq 5/2$.
With this, the tightness of the decomposition inner bound is proved.
The example highlights the key steps for the proof of optimality of the lower bound in~\eqref{eq:sumDoF in} for the general $K$-user case without a dominant user.

\subsection{Channel transformation}
\label{sec:example:trans}

In general, we can set $\bar{X}_i = \mathbf{V}_i {X}_i$ and construct ${Y}_i = \mathbf{U}_i \bar{Y}_i$  in the channel in~\eqref{eq:ch model general}, where the beamforming matrices $\mathbf{V}_i$ and the shaping matrices $\mathbf{U}_i$ are full-rank / invertible square matrices of dimension $M_i$, $i\in[1:K]$ that do not depend on $\mathsf{P}$. Since invertible transformations preserve DoF, the channel in~\eqref{eq:ch model general} and the transformed one have the same DoF.
The input-output relationship of the transformed channel reads
\begin{subequations} 
\begin{align}
{Y}_i &= \sum_{k\in[1:K]} H_{ik} {X}_k + Z_i\in\mathbb{C}^{M_i \times 1},  
\label{eq:ch model general transformed}
\\
H_{ik} &:= \mathbf{U}_i \bar{H}_{ik}\mathbf{V}_k\in\mathbb{C}^{M_i \times M_k}, \  (i,k)\in[1:K]^2,
\end{align} 
\end{subequations} 
where in~\eqref{eq:ch model general transformed} we neither specify the input power constraints on the inputs $ {X}_k, \ k\in[1:K],$ nor the covariance matrix of the noise terms, as they do not impact the DoF.

In the following we assume that all channel coefficients are {\em generic}, i.e., randomly chosen from a continuous distribution
. Under this assumption, the goal is to show how to find {\it invertible} beamforming and shaping matrices such that the transformed channel for our $(M_1,M_2,M_3)=(2,2,1)$ example is
\begin{align*}
Y_{1} & =
\begin{bmatrix}
h^{(11)}_{11} & h^{(11)}_{12} \\
h^{(11)}_{21} & h^{(11)}_{22} \\
\end{bmatrix} X_1+ 
\begin{bmatrix}
h^{(12)}_{11} & 0 \\
0 & h^{(12)}_{22} \\
\end{bmatrix} X_2+
\begin{bmatrix}
h^{(13)}_{11} \\
0 \\
\end{bmatrix} X_{3},
\\
Y_{2} & = 
\begin{bmatrix}
h^{(21)}_{11} & 0 \\
0 & h^{(21)}_{22} \\
\end{bmatrix} X_1+ 
\begin{bmatrix}
h^{(22)}_{11} & h^{(22)}_{12} \\
h^{(22)}_{21} & h^{(22)}_{22} \\
\end{bmatrix} X_2+ 
\begin{bmatrix}
h^{(23)}_{11} \\
0 \\
\end{bmatrix} X_{3},
\\
Y_{3} & = 
\begin{bmatrix}
h^{(31)}_{11} &
0  \\
\end{bmatrix} X_1+ 
\begin{bmatrix}
h^{(32)}_{11} &
0 \\
\end{bmatrix} X_2+ 
h^{(33)}_{11} X_{3},
\end{align*}
where $h_{ab}^{(ij)}$ is the scalar channel gain from the $b$-th antenna of ${\rm Tx}_j$ to the $a$-th antenna of ${\rm Rx}_i$, and where we no longer write the noises for notation convenience.
To show that indeed such a transformed channel can be found, we proceed along a number of steps.

\paragraph*{Step~1}
As a first step we neutralize at (the single antenna of) ${\rm Rx}_3$ the signal from the second antenna of ${\rm Tx}_1$ and from the second antenna of ${\rm Tx}_2$. We do so by carefully choosing some columns of the matrices $\mathbf{V}_1$ and $\mathbf{V}_2$. 
Let
\begin{align*}
\mathbf{V}_1 := 
\begin{bmatrix} v_{11} & v_{12} \end{bmatrix},
\
\mathbf{V}_2 := 
\begin{bmatrix} v_{21} & v_{22} \end{bmatrix},
\end{align*}
where $v_{ki}$ indicates the $i$-th column of the matrix $\mathbf{V}_k$.
We choose $(v_{12},v_{22})$ such that
\begin{align*}
\bar{H}_{31}v_{12} = 0, \
\bar{H}_{32}v_{22} = 0.
\end{align*}
Since $\bar{H}_{32}$ and $\bar{H}_{31}$ are generic $1\times2$ matrices,
$v_{12}$ and $v_{21}$ (which are $2\times1$ matrices) can be chosen from the (one dimensional) right null space of $\bar{H}_{32}$ and $\bar{H}_{31}$, respectively.

\paragraph*{Step~2}
As a second step we neutralize the signal 
from the second antenna of ${\rm Tx}_1$ 
at the first antenna of ${\rm Rx}_2$, 
and from the second antenna of ${\rm Tx}_2$ 
at the first antenna of ${\rm Rx}_1$. 
We let
\begin{align*}
\mathbf{U}_1 := \begin{bmatrix}
u_{11}\\
u_{12}\\
\end{bmatrix}, \
\mathbf{U}_2 := \begin{bmatrix}
u_{21}\\
u_{22}\\
\end{bmatrix},
\end{align*}
where $u_{ki}$ indicates the $i$-th row of the matrix $\mathbf{U}_k$.
In order to achieve our goal, we impose 
\begin{align*}
u_{11}\bar{H}_{12}v_{22} = 0, \
u_{21}\bar{H}_{21}v_{12} = 0.
\end{align*}
Since $v_{12}$ and $v_{22}$ have been decided already based on $\bar{H}_{31}$ and $\bar{H}_{32}$, we have that $\bar{H}_{12}v_{22}$ and $\bar{H}_{21}v_{12}$ are generic $2\times1$ matrices almost surely.
Therefore, $u_{11}$ and $u_{21}$ (which are $1\times2$ matrices) can be chosen from the (one dimensional) left null space of $\bar{H}_{12}v_{22}$ and $\bar{H}_{21}v_{12}$, respectively.

\paragraph*{Step~3}
As a third step, we neutralize the signal from (the single antenna of) ${\rm Tx}_3$ 
at the second antenna of ${\rm Rx}_1$ and at the second antenna of ${\rm Rx}_2$. 
We thus impose
\begin{align*}
u_{12}\bar{H}_{13}\mathbf{V}_3 = 0, \
u_{22}\bar{H}_{23}\mathbf{V}_3 = 0.
\end{align*}
Since $\mathbf{V}_3$ is a non-zero scalar, we choose $u_{12}$ and $u_{22}$ as rows in the (one dimensional) left null space of $\bar{H}_{13}$ and $\bar{H}_{23}$, respectively.

\paragraph*{Step~4}
As a last step, we neutralize the signal received at the second antenna of ${\rm Rx}_2$ 
from the first antenna of ${\rm Tx}_1$ 
and the one received at the second antenna of ${\rm Rx}_1$ 
from the first antenna of ${\rm Tx}_2$. 
For this we impose
\begin{align*}
u_{12}\bar{H}_{12}v_{21} = 0, \
u_{22}\bar{H}_{21}v_{11} = 0.
\end{align*}
Since $u_{12}$ and $u_{22}$ have been decided already based on $\bar{H}_{13}$ and $\bar{H}_{23}$,
the vectors $u_{12}\bar{H}_{12}$ and $u_{22}\bar{H}_{21}$ have dimension $1\times2$ and are generic.
Therefore, we choose $v_{21}$ and $v_{11}$ to be columns in their respective (one dimensional) right null spaces.

\medskip
By the above operations, $\mathbf{V}_1,\mathbf{V}_2,\mathbf{U}_1,\mathbf{U}_2$ have been decided.
$\mathbf{V}_3$ and $\mathbf{U}_3$ are scalars and can be set to one without loss of generality.
Also all transform matrices were decided based on generic channel coefficients, so they do not have dependence or special structure.
Thus all transform matrices are full rank and invertible almost surely, and the transformation preserves the DoF.

\subsection{An intuitive dimension-counting argument}
\label{sec:example:outLin}
We start with a `dimension counting' argument to give an intuitive reason as to why the decomposition inner bound $d_\Sigma \geq 5/2$ should be tight. Without loss of generality, we can assume $d_1=d_2=d$ and $d_3=d^{\prime}$.

Since ${\rm Rx}_3$ has a single antenna, the total DoF of its own and the interference signal cannot be larger than one.
This implies that the interference at ${\rm Rx}_3$ must have less than $1-d^{\prime}$~DoF. 

Now consider ${\rm Tx}_2$ that must achieve $d$~DoF.
Since the part of its signal that causes interference at ${\rm Rx}_3$ must have less than $1-d^{\prime}$~DoF,
the part of its signal that does not interfere at ${\rm Rx}_3$ must have at least $d-\left(1-d^{\prime}\right)$~DoF.
This is to say, ${\rm Tx}_2$ controls $d-\left(1-d^{\prime}\right)$ dimensions to be neutralized at ${\rm Rx}_3$ and these dimensions are therefore decided.

Now consider ${\rm Rx}_1$, which has two antennas. 
By the generic setting, 
the decided $d-\left(1-d^{\prime}\right)$
dimensions at ${\rm Tx}_2$ do not align automatically with the
interference from ${\rm Tx}_3$; therefore we have the bound 
\begin{subequations}
\begin{align}
[d]+
[d-(1-d^{\prime})]+
[d^{\prime}]
\leq2
\Longleftrightarrow
d+d^{\prime}\leq\frac{3}{2}.
\end{align}

From the outer bound in~\eqref{eq:sumDoF out} and from $d_i\leq M_i, \ i\in[1:3]$, we know
\begin{align}
2d \leq  2, \ d^{\prime}  \leq  1.
\end{align}
\label{eq: new bound example}
\end{subequations}
It is easy to see that the bounds in~\eqref{eq: new bound example} define a pentagon with vertices $(d,d^{\prime})=(1,1/2)$ and $(d,d^{\prime})=(1/2,1)$. Therefore the largest DoF can be at most $2d+d^{\prime} = 5/2$.
Since a DoF of $\frac{5}{2}$ is achievable by~\eqref{eq:sumDoF in}, we conclude that the cooperation outer bound in~\eqref{eq:sumDoF out} might be loose.

\subsection{An information theoretic proof}
\label{sec:example:outIT}

We define the differential entropy of the noisy signal as in \cite{noisy-differential-entropy}
\begin{align}
  \hbar(X^n):=h(X^n+Z^n),
\end{align}
 where $h$ is standard differential entropy. 
 $X^n$ is a signal vector power constrained to $\mathsf{P}$ and $Z^n\sim \mathcal{N}(0,I)$ is independent noise vector. 
 Joint and conditional differential entropies are defined in the same manner \cite{noisy-differential-entropy}.

We next formalize the intuitive argument from Section~\ref{sec:example:outLin}.
In the transformed channel, 
let $X_{k1}$ be the signal sent by the first antenna of ${\rm Tx}_k$, 
and $X_{k2}$ be one sent by the second antenna of ${\rm Tx}_k$, $k\in[1:2]$. 
By Fano's inequality, we have 
\begin{subequations}
\begin{align}
n&(R_3 -\epsilon_n) 
\\&\leq I(W_3;Y_3^n) 
   =  h\left(Y^n_{3}\right)-h\left(Y_{3}^n|W_{3}\right)
\\&\leq h\left(Y^n_{3}\right)-h\left(Y_{3}^n|W_{3},W_{1}\right)
\\&\leq n\left(1 \cdot \log(\mathsf{P})+o(\log(\mathsf{P}))\right)-h \left(Y_{3}^n|W_{3},W_{1}\right) \label{eq:one_antenna}
\\& = n\left(1 \cdot \log(\mathsf{P})+o(\log(\mathsf{P}))\right)-\hbar \left(X_{21}^n\right),
\label{eq:reduce_side_info}
\end{align}
\label{eq: new bound rig 1}
\end{subequations}
where
the inequality in~\eqref{eq:one_antenna} follows because ${\rm Rx}_3$ has only one antenna,
and the one in~\eqref{eq:reduce_side_info} since in the transformed channel 
\begin{align*}
&h\Big(Y_{3}^n|W_{3},W_{1}\Big)
\\&=h\Big(h^{(31)}_{11} X_{11}^n+h^{(32)}_{11} X_{21}^n+ h^{(33)}_{11} X_{3}^n+Z_{3}^n|W_{3},W_{1}\Big)
\\&=h\Big(h^{(32)}_{11} X_{21}^n+Z_3^n\Big)=:\hbar(X_{21}^n),
\end{align*}
which implies that ${\rm Rx}_3$ can recover $X_{21}^n$  up to noise distortion of the order $o(\log(\mathsf{P}))$.
Hence, the bound in~\eqref{eq: new bound rig 1} implies 
\begin{align}
\hbar\left(X_{21}^n\right)\leq n\left(1 \cdot \log(\mathsf{P})-R_3+\epsilon_n+o(\log(\mathsf{P}))\right).
\label{eq: h x2a 1}
\end{align}
Moreover, the bound in~\eqref{eq: h x2a 1} together with 
\begin{align*}
n&(R_2 -\epsilon_n)\leq I(W_2;Y_2^n) \leq I(X_2^n;Y_2^n)
\\&=\hbar\left(X_{21}^n,X_{22}^n\right)=\hbar(X_{21}^n)+\hbar\left(X_{22}^n|X_{21}^n\right),
\end{align*}
implies
\begin{align}
\hbar\left(X_{22}^n|X_{21}^n\right)\geq n\left(R_2- 1 \cdot \log(\mathsf{P})+R_3-2\epsilon_n+o(\log(\mathsf{P}))\right).
\label{eq:ineqn}
\end{align}
Now consider
\begin{subequations}
\begin{align}
n& (R_1-\epsilon_n)
 \leq  I\left(Y^n_{1};W_{1}\right) 
\\& \leq  h\left(Y^n_{1}\right)-h\left(Y^n_{1}|W_{1},X^n_{21}\right)\label{eq:condition_2}
\\& \leq  
n\left(2 \cdot\log(\mathsf{P})+o(\log(\mathsf{P}))\right)-\hbar\left(X^n_{22},X^n_3|X^n_{1},X^n_{21}\right) \label{eq:2_antenna}
\\& = n\left(2 \cdot\log(\mathsf{P})+o(\log(\mathsf{P}))\right)-\hbar\left(X^n_{3}\right)-\hbar\left(X^n_{22}|X^n_{21}\right)\label{eq:independent}
\\& \leq  n ( 3 \cdot \log(\mathsf{P}) - 2R_3-R_2+3\epsilon_n+o(\log(\mathsf{P}))), \label{eq:final_step}
\end{align}
\label{eq: new bound rig 2}%
\end{subequations}
where the inequality in~\eqref{eq:2_antenna} follows since ${\rm Rx}_1$ has two antennas, 
the one in~\eqref{eq:independent} since $X_{3}^n$ and $X_{2}^n=(X_{21}^n,X_{22}^n)$ are independent,
and finally~\eqref{eq:final_step} comes from \eqref{eq:ineqn} and $n(R_3-\epsilon_n)\leq  I\left(Y^n_{3};X_{3}^n\right) \leq \hbar(X_3^n)$. 

Therefore, from~\eqref{eq: new bound rig 2} and for $n\gg1$, we conclude that 
\begin{align}
\frac{R_1+R_2+2R_3}{\log(\mathsf{P})} \leq 3 +o(1),
\end{align}
or equivalently that $d+d^{\prime} \leq 3/2$ (recall $R_1=R_2=d\log(\mathsf{P})$ and $R_3=d^{\prime}\log(\mathsf{P})$ without loss of optimality for DoF). 

The argument at the end of Section~\ref{sec:example:outLin} shows that the novel bound $d+d^{\prime} \leq 3/2$, together with known outer bounds, implies $d_\Sigma \leq 5/2$. Since the outer bound is achievable by the decomposition inner bound, we have $d_\Sigma = 5/2$. This completes the proof for this specific example.

\section{Sum DoF for the general $K$-user case}
\label{sec:Kuser}

In the previous section, through suitable invertible transformations we could rewrite the original IC into a new one with a special structure in the channel matrices; this structure suggested how to provide genie side information to the receivers in the outer bound proof.
We extend here the proof for the example in Section~\ref{sec:example} in two ways. 
First we give a DoF outer bound for the general 3-user IC with number of antennas specified by the vector $(M_1,M_2,M_3)$ in Section~\ref{sec:Kuser:K=3}. Then we generalize the result to the $K$-user case in Section~\ref{sec:Kuser:Kgen}.

\subsection{The 3-user case}
\label{sec:Kuser:K=3}
Without loss of generality let $M_1\geq M_2\geq M_3$. We assume there is no dominant user, that is, $M_1< M_2+M_3$.
By the invertible transformations $\bar{X}_i = \mathbf{V}_i {X}_i, \ {Y}_i = \mathbf{U}_i \bar{Y}_i, \ i\in[1:3],$
we aim to obtain an equivalent channel where the inputs are partitioned as
${X}_1=(X_{11},X_{12},X_{13})$,
${X}_2=(X_{21},X_{22})$, and 
${X}_3=(X_{31},X_{32})$, and 
similarly for the outputs. 
Let $|X_{ij}|$ indicate the size / number of antennas in $X_{ij}$.
We want
\begin{align*}
  |X_{11}| = |Y_{13}| = |X_{32}| = |Y_{32}| &= M_1-M_2,\nonumber\\  
  |X_{13}| = |Y_{11}| = |X_{22}| = |Y_{22}| &= M_1-M_3,\nonumber\\
  |X_{12}| = |Y_{12}| = |X_{21}| = |Y_{21}| \nonumber\\= |X_{31}| = |Y_{31}| &= M_2+M_3-M_1.\nonumber
\end{align*}
Let the channel matrix from $\bar{X}_{jb}$ to $\bar{Y}_{ia}$ in the original channel be denoted as $\bar{h}^{(ij)}_{ab}$, with size $|Y_{ia}|\times |X_{jb}|$.

We now derive the channel input/output relationship of the transformed channel.
As before, the beamforming matrices in \eqref{eq:ch model general transformed} are denoted as
\begin{align*}
\mathbf{V}_1 = 
\begin{bmatrix} v_{11} & v_{12} & v_{13} \end{bmatrix},
\
\mathbf{V}_2 = 
\begin{bmatrix} v_{21} & v_{22} \end{bmatrix},
\
\mathbf{V}_3 =
\begin{bmatrix} v_{31} & v_{32} \end{bmatrix},
\end{align*}
where $v_{ab}$ has size $M_a \times |X_{ab}|$, and the shaping matrices as 
\begin{align*}
\mathbf{U}_1 = \begin{bmatrix}
u_{11}\\
u_{12}\\
u_{13}
\end{bmatrix}, \
\mathbf{U}_2 = \begin{bmatrix}
u_{21}\\
u_{22}\\
\end{bmatrix}, \
\mathbf{U}_3 = \begin{bmatrix}
u_{31}\\
u_{32}\\
\end{bmatrix},
\end{align*}
where $u_{ab}$ has size $|Y_{ab}| \times M_a $.

We first choose the beamforming matrices by imposing
\begin{align*}
  \bar{H}_{21}v_{11}=0,\ \
  \bar{H}_{31}v_{13}=0,\\
  \begin{bmatrix}
    \bar{h}^{(12)}_{11} & \bar{h}^{(12)}_{12}  \\
  \end{bmatrix}
  v_{21}=0, \\
  \begin{bmatrix}
    \bar{h}^{(13)}_{31} & \bar{h}^{(13)}_{32}  \\
  \end{bmatrix}
  v_{31}=0. 
\end{align*}
Under the generic channel gain assumption, the matrices 
$\bar{H}_{21}$, $\bar{H}_{31}$, 
$\begin{bmatrix} \bar{h}^{(12)}_{11} & \bar{h}^{(12)}_{12}  \\ \end{bmatrix}$ and 
$\begin{bmatrix} \bar{h}^{(13)}_{31} & \bar{h}^{(13)}_{32}  \\ \end{bmatrix}$ 
have right null space of rank $M_1-M_2$, $M_1-M_3$, $M_2+M_3-M_1$, and $M_2+M_3-M_1$, respectively, almost surely. 
Thus we can pick columns from these right null spaces to form the beamforming matrices
$v_{11},\ v_{13},\ v_{21},\ v_{31}$, which are therefore of size  $M_1\times (M_1-M_2)$, $M_1\times (M_1-M_3)$, $M_2\times (M_2+M_3-M_1)$, and $M_3 \times (M_2+M_3-M_1)$, respectively, and are still generic almost surely. 
The matrices $v_{12}$, $v_{22}$, and $v_{32}$ are randomly chosen so that they are full-rank and with no specific relation with the previously chosen matrices.

We then choose the shaping matrices by imposing
\begin{align*}
  u_{13}\bar{H}_{12} = 0, \ \
  u_{11}\bar{H}_{13} = 0, \\
  u_{12} 
  \begin{bmatrix}
    \bar{h}^{(12)}_{12} & \bar{h}^{(13)}_{12}  \\
    \bar{h}^{(12)}_{22} & \bar{h}^{(13)}_{22}  \\
    \bar{h}^{(12)}_{32} & \bar{h}^{(13)}_{32}  \\
  \end{bmatrix} = 0,\\
  u_{21}
  \begin{bmatrix}
    \bar{h}^{(21)}_{13} \\ \bar{h}^{(21)}_{23}  \\
  \end{bmatrix} =0,
  \ \
  u_{22} \begin{bmatrix}
    \bar{h}^{(23)}_{11} \\ \bar{h}^{(23)}_{21}  \\
  \end{bmatrix} =0,\\
  u_{31} \begin{bmatrix}
    \bar{h}^{(31)}_{11} \\ \bar{h}^{(31)}_{21}  \\
  \end{bmatrix} =0,
  \ \
  u_{32} \begin{bmatrix}
    \bar{h}^{(32)}_{11} \\ \bar{h}^{(32)}_{21}  \\
  \end{bmatrix} =0.
\end{align*}
Under the generic channel gain assumption, all channel matrices are full rank almost surely;
the shaping matrices can thus be chosen as rows is the respective right null spaces and are still generic almost surely.
$U_2$ is full rank matrix, since $u_{21}$ and $u_{22}$ are chosen from independent null spaces, thus are independent.  Similarly, we claim $U_3$ is full rank. We then show that $U_1$ is also full rank. It is easy to see that $u_{11}$ and $u_{13}$ are independent. If $U_1$ is not full rank, there must exist non-zero row-vectors $g_1$ of size $1\times (2M_1-M_2-M_3)$ and $g_2$ of size $1\times (M_2+M_3-M_1)$ such that 
\begin{align*}  
  g_1 \begin{bmatrix}
    u_{11}\\
    u_{13}
  \end{bmatrix}
  =
g_2 u_{12},
\end{align*}  
that is 
\begin{align*}
  \mathbf{0} &= g_2 u_{12} \begin{bmatrix}
    \bar{h}^{(12)}_{12} & \bar{h}^{(13)}_{12}  \\
    \bar{h}^{(12)}_{22} & \bar{h}^{(13)}_{22}  \\
    \bar{h}^{(12)}_{32} & \bar{h}^{(13)}_{32}  \\
  \end{bmatrix}\\
  &= g_1 \begin{bmatrix}
    u_{11}\\
    u_{13}
  \end{bmatrix}
  \begin{bmatrix}
    \bar{h}^{(12)}_{12} & \bar{h}^{(13)}_{12}  \\
    \bar{h}^{(12)}_{22} & \bar{h}^{(13)}_{22}  \\
    \bar{h}^{(12)}_{32} & \bar{h}^{(13)}_{32}  \\
  \end{bmatrix}\\
  &= g_1 
  \begin{bmatrix}
    u_{11}
    \begin{bmatrix}
     \bar{h}^{(12)}_{12}  \\
     \bar{h}^{(12)}_{22}  \\
     \bar{h}^{(12)}_{32}  \\
  \end{bmatrix} & \mathbf{0} \\
     \mathbf{0} & u_{13}
    \begin{bmatrix}
    \bar{h}^{(13)}_{12} \\
    \bar{h}^{(13)}_{22} \\
    \bar{h}^{(13)}_{32} \\
  \end{bmatrix} \\
  \end{bmatrix}\\
  &= g_1 \mathbf{F}.
\end{align*}
Since $u_{11}$ is independent of $\bar{H}_{12}$ and $u_{13}$ is independent of $\bar{H}_{13}$, $\mathbf{F}$ is a full rank square matrix almost surely. Then $G_1=\mathbf{0}$, which contradicts our initial assumption. Therefore we claim $U_1$ is full rank almost surely.

With the chosen beamforming and shaping matrices, the transformed channel has input/output relationship 
\begin{align*}
Y_{1}  =&
\begin{bmatrix}
h^{(11)}_{11} & h^{(11)}_{12} & h^{(11)}_{13}\\
h^{(11)}_{21} & h^{(11)}_{22} & h^{(11)}_{23}\\
h^{(11)}_{31} & h^{(11)}_{32} & h^{(11)}_{33}\\
\end{bmatrix} X_1+ 
\begin{bmatrix}
0 & h^{(12)}_{12} \\
h^{(12)}_{21} & 0 \\
0 & 0 \\
\end{bmatrix} X_2\nonumber\\
& +\begin{bmatrix}
0 & 0 \\
h^{(13)}_{21} & 0 \\
0 & h^{(13)}_{32} \\
\end{bmatrix} X_{3}
\\
Y_{2}  =& 
\begin{bmatrix}
0 & h^{(21)}_{12} & 0 \\
0 & h^{(21)}_{22} & h^{(21)}_{23} \\
\end{bmatrix} X_1+ 
\begin{bmatrix}
h^{(22)}_{11} & h^{(22)}_{12} \\
h^{(22)}_{21} & h^{(22)}_{22} \\
\end{bmatrix} X_2 \nonumber\\
&+\begin{bmatrix}
h^{(23)}_{11} & h^{(23)}_{12}\\
0 & h^{(23)}_{22} \\
\end{bmatrix} X_{3}
\\
Y_{3}  = &
\begin{bmatrix}
0 & h^{(31)}_{12} & 0 \\
h^{(31)}_{21} & h^{(31)}_{22} & 0 \\
\end{bmatrix} X_1+ 
\begin{bmatrix}
h^{(32)}_{11} & h^{(32)}_{12} \\
0 & h^{(32)}_{22} \\
\end{bmatrix} X_2\nonumber\\
&+\begin{bmatrix}
h^{(33)}_{11} & h^{(33)}_{12} \\
h^{(33)}_{21} & h^{(33)}_{22} \\
\end{bmatrix} X_{3},
\end{align*}
where $h^{(ij)}_{ab}$ represents the transformed channel matrix from $X_{jb}$ to $Y_{ia}$, which has size $|Y_{ia}|\times |X_{jb}|$. Since the beamforming and shaping matrices are full rank almost surely, we performed an invertible transformation that preserves the DoF. Therefore we obtained a new channel that is not fully connected and whose structure suggests which genie side information to provide to the receivers in the converse proof. 

We shall consider different choices of side information at the various receivers.
The idea is to start as usual by Fano's inequality, by providing side information $S_u^n$ to receiver $u$, and by bounding the entropy of the output as a function of the number of antennas at receiver $u$, so as to obtain
\begin{align*}
  n&(R_u-\varepsilon_n)  \leq n(M_u \log(\mathsf{P})+o(\log(\mathsf{P})))-h(Y_u^n|W_u,S_u^n).
\end{align*}
The entropy term $h(Y_u^n|W_u,S_u)$ depends on the distribution of the interference at receiver $u$ (since $X_u^n$ can be cancelled thanks to the knowledge of $W_u$) conditioned on the side information $S_u^n$; if such an entropy term, which appears with a negative sign, cannot be single-letterized, then we proceed to provide side information to another receiver in such a way that the same entropy term appears with positive sign; by adding the two bounds we `get rid' of the entropy terms that cannot be single-letterized. We continue in this fashion until we obtain a single-letter outer bound.
For the general asymmetric 3-user MIMO IC the steps are as follows.

\paragraph*{1st bound: message side information}
By providing ${\rm Rx}_2$ with side information $W_3$ we have
\begin{align}
  n&(R_2-\varepsilon_n)  \leq h(Y_2^n)-h(Y_2^n|W_2,W_3)\nonumber\\
  & \leq n(M_2 \log(\mathsf{P})+o(\log(\mathsf{P})))-\hbar(X^n_{12},X^n_{13}|X_2^n,X_3^n) \nonumber\\ 
  &= n(M_2 \log(\mathsf{P})+o(\log(\mathsf{P})))-\hbar(X^n_{12},X^n_{13}),
  \label{eq:1}
\end{align}
where the inequality 
follows since ${\rm Rx}_2$ does not receive $X_{11}$. 
Similarly, by providing ${\rm Rx}_3$ with $W_2$ we obtain
\begin{align}
  n(R_3-\varepsilon_n) & \leq n(M_3 \log(\mathsf{P})+o(\log(\mathsf{P})))
 -\hbar(X_{11}^n,X_{12}^n) \label{eq:2}.
\end{align}
By adding~\eqref{eq:1} and~\eqref{eq:2} and since
\begin{align*}
&
 \hbar(X_{11}^n,X_{12}^n)
+\hbar(X_{12}^n,X_{13}^n)
\\&\geq \hbar(X_{11}^n,X_{12}^n)
+\hbar(X_{13}^n|X_{12}^n,X_{11}^n)+\hbar(X_{12}^n)
\\&=  \hbar(X_{11}^n,X_{12}^n,X^n_{13})+\hbar(X_{12}^n)
\\&\geq n(R_1-\varepsilon_n)+\hbar(X_{12}^n),
\end{align*}
we obtain
\begin{align}
  n&(R_1+R_2+R_3-3\varepsilon_n) \nonumber
  \\&\leq n((M_2+M_3)\log(\mathsf{P})+o(\log(\mathsf{P}))) -\hbar(X_{12}^n).
  \label{eq:1st}
\end{align}

\paragraph*{2nd bound: ``signal pieces'' side information}
Next, we provide $(X_{12}^n,X_{32}^n)$ as side information to ${\rm Rx}_2$ and obtain
\begin{align}
  n&(R_2-\varepsilon_n)  \leq h(Y_2^n)-h(Y_2^n|W_2,X_{12}^n,X_{32}^n)\nonumber\\
  & \leq n(M_2 \log(\mathsf{P})+o(\log(\mathsf{P}))) -\hbar(X_{31}^n,X_{11}^n |W_2,X_{12}^n,X_{32}^n)\nonumber\\
  &= n(M_2 \log(\mathsf{P})+o(\log(\mathsf{P})))-\hbar(X_{31}^n|X_{32}^n)-\hbar(X_{13}^n|X_{12}^n).
  \label{eq:3}
\end{align}
Similarly, we provide $(X_{11}^n,X_{21}^n)$ as side information to ${\rm Rx}_3$ and obtain
\begin{align}
  n(R_3-\varepsilon_n) & \leq n(M_3 \log(\mathsf{P})+o(\log(\mathsf{P})))\nonumber\\
  &\quad -\hbar(X_{21}^n|X_{22}^n)-\hbar(X_{11}^n|X_{12}^n). \label{eq:4}
\end{align}
By adding \eqref{eq:3} and \eqref{eq:4} we obtain
\begin{align}
  n&(R_1+R_2+R_3-3\varepsilon_n) \leq
  n((M_2+M_3)\log(\mathsf{P})\nonumber\\
  &+o(\log(\mathsf{P})))+\hbar(X_{12}^n)-\hbar(X_{21}^n|X_{22}^n)-\hbar(X_{31}^n|X_{32}^n).
  \label{eq:2nd}
\end{align}

\paragraph*{3rd bound: MAC bounds}
Now, we provide ${\rm Rx}_1$ with enough side information to enable the decoding of all messages.
After ${\rm Rx}_1$ has decoded its own message / removed $X_1^n$ from the received signal, it is left with $M_1$ linear combinations of $M_2+M_3$ interfering symbols; if we we provide ${\rm Rx}_1$ with $M_2+M_3-M_1$ extra observations / antenna outputs, it will be able to decode all interfering symbols. Next we derive two such `MAC-bounds' by providing either $X_{21}^n$ or $X_{31}^n$ to ${\rm Rx}_1$. We have
\begin{align}
  n&(R_1+R_2+R_3-3\varepsilon_n) \leq I(W_1,W_2,W_3;Y_1^n,X_{21}^n)\nonumber\\
  &=     h(Y_1^n,X_{21}^n)-o(\log(\mathsf{P})) \nonumber\\
  &\leq  h(Y_1^n)+\hbar(X_{21}^n|X_{22}^n)-o(\log(\mathsf{P}))\nonumber\\
  &=     h(Y_1^n)+\hbar(X_{21}^n, X_{22}^n)-\hbar(X_{22}^n)-o(\log(\mathsf{P}))\nonumber\\
  &\leq  n(M_1 \log(\mathsf{P})+o(\log(\mathsf{P})))+n(R_2-\varepsilon_n)-\hbar(X_{22}^n).
  \label{eq:3rd}
\end{align}
Similarly
\begin{align}
  n&(R_1+R_2+R_3-3\varepsilon_n) \leq I(W_1,W_2,W_3;Y_1^n,X_{31}^n)\nonumber\\
  &\leq n(M_1 \log(\mathsf{P})+o(\log(\mathsf{P})))+n(R_3-\varepsilon_n)-\hbar(X_{32}^n).
  \label{eq:4th}
\end{align}

\paragraph*{Final bound}
By adding \eqref{eq:1st}, \eqref{eq:2nd}, \eqref{eq:3rd}, \eqref{eq:4th}, and by taking $n\rightarrow \infty$, we obtain
\begin{align*}
  4R_1+4R_2+4R_3 &\leq 2(M_1+M_2+M_3) \log(\mathsf{P})+o(\log(\mathsf{P})),
\end{align*}
and therefore the DoF is outer bounded by 
\begin{align*}
  d_\Sigma 
  &\leq  \lim_{P\rightarrow \infty}  \frac{2(M_1+M_2+M_3)\log(\mathsf{P})+o(\log(\mathsf{P}))}{4\log{(\mathsf{P})}}\nonumber\\ 
  &= \frac{M_1+M_2+M_3}{2}.
\end{align*}
This concludes the proof for the general 3-user asymmetric IC in the case where there is no dominant user.

\subsection{The general $K$-user case}\label{sec:Kuser:Kgen}
We are now ready to extend our 3-user result to the general $K$-user asymmetric MIMO IC.
Our main result is 
\begin{theorem}
\label{thm:k_user_sum_dof}
For almost all channel realizations  the asymmetric $K$-user MIMO IC, in which the $i$-th user has $M_i$ antennas at both the transmitter and the receiver, $i\in[1:K]$, the DoF is
\begin{align}
  d_\Sigma=\max \left(\sum_{i\in [1:K]}M_i/2, \ \max_{i\in [1:K]}M_i \right).
\end{align}
\end{theorem}

\begin{IEEEproof}
As per our discussion in Section~\ref{sec:model:dom}, when there is a dominant user (whose has more antennas than the rest of the users combined) it is optional to let only that user transmit. 
When there is no dominant user, we can always partition the users into three groups such that no group has more antennas than the the other two groups combined. Then we allow the users in the same group to fully cooperate and apply our bound for 3-user IC, which shows that the DoF is half the sum of number of the total number of antennas. 
This concludes the proof.
\end{IEEEproof}

\section{Conclusion}\label{sec:conclusion}
In this paper we studied a special class of $K$-user asymmetric MIMO interference channels in which a transmitter and its receiver are equipped with the same number of antennas, while different users may have different number of antennas. 
We showed that existing cooperation-based outer bounds are loose and gave a novel outer bound. 
Our result indicates two operating regimes.
For systems with a dominant user (a user who has more antennas that the other users combined), the optimal DoF is achieved by inactivating all but the dominant user.
For systems without a dominant user, the decomposition inner bound turns out to be tight, that is, the MIMO operations do not help in the DoF perspective.
The characterization of the DoF of arbitrary asymmetric $K$-user MIMO interference channels is part of ongoing investigation.

\section*{Acknowledgement}\label{sec:acknowledgement}
The work of T. Liu and D.~Tuninetti was partially funded by NSF under award number 1218635. S. Jafar is affiliated with the Center for Pervasive Communications and Computing (CPCC) at UC Irvine. His work is supported in part by NSF grant CCF-1319104. The contents of this article are solely the responsibility of the author and do not necessarily represent the official views of the NSF.

\baselineskip=2pt
\bibliographystyle{ieeetr}
\bibliography{refs}

\end{document}